# Pareto-optimal alloys

Thomas Bligaard, Gisli H. Johannesson, Andrei V. Ruban, Hans L. Skriver, Karsten W. Jacobsen, and Jens K. Nørskov

*Center for Atomic-scale Materials Physics, Department of Physics, Technical University of Denmark, DK-2800 Lyngby, Denmark.*

**Large databases that can be used in the search for new materials with specific properties remain an elusive goal in materials science. The search problem is complicated by the fact that the optimal material for a given application is usually a compromise between a number of materials properties and the price. In this letter we present a database consisting of the lattice parameters, bulk moduli, and heats of formation for over 64,000 ordered metallic alloys, which has been established by direct first-principles density-functional-theory calculations. Furthermore, we use a concept from economic theory, the Pareto-optimal set, to determine optimal alloy solutions for the compromise between low compressibility, high stability and price.**

It would be extremely valuable if one could establish a "materials informatics" approach in searching for new materials. If ideas for new materials could be generated by suitable searches in databases, one could decrease the number of expensive experiments that need to be done. One approach in this direction has been the "Materials Selector" introduced by Ashby[1]. Here selection of a material for a particular application is made on the basis of a database of *existing* materials. When it comes to the prediction of *new* materials including compositions and structures that have not yet been synthesized and tested there are presently very few approaches. One simple example is the very successful "Miedema model[2]" which relates alloy heats of formation to two parameters characterizing each metallic element in the periodic table.

The underlying problem in any materials informatics approach today is that even though a number of materials properties are now assembled, there are still very few experimental data compared to the vast number of possible material combinations. One way of increasing the amount of data in materials science is to rely on first principles density functional calculations. The calculations now have a level of accuracy, which is often comparable to experiments, in particular when the aim is to describe variations from one system to the next[3-5]. We have used our density functional calculations to evaluate the equation of state (the relation between energy and volume of the material) for 64,149 different alloys with up to four components.

For each alloy we perform first-principles total-energy calculations for at least 4 different unit cell volumes and, subsequently, fit the energy-volume curve to an equation of state. The calculations have been performed within an LMTO-ASA implementation[6] of the generalised gradient approximation[7] for the exchange and correlation energy (see reference 5 for details). A database of the derived materials parameters: lattice constants, heats of formation, and bulk moduli for all the calculated alloys is available at http://www.fysik.dtu.dk/CAMPMDB/QuarternaryAlloys/

The database contains 64,149 out of a total of 192,016 possible four (or less)-component alloys in the face-centred cubic and body-centred cubic structures with four atom unit cells and, therefore, we cannot be sure that all interesting alloys are included. However, the database was made in a search for the most stable alloys[5] and since these are also the alloys that may most likely be synthesized and remain unchanged afterwards, the database should contain some of the most interesting alloys from an application point of view.

Given the database, the question is how one may use it to search for new materials. We will illustrate this by searching for the alloys that have the smallest compressibility. A low compressibility by itself is interesting, since it means that the material is very hard[8], but it is also interesting from other points of view. It has, for instance, been shown (page 52 in reference 1) that there is a monotonic relationship between the Young's modulus and the thermal expansion coefficient, and the Young's modulus is approximately equal to the inverse compressibility for most metals. A low compressibility is therefore a good sign of a low thermal expansion coefficient, something that is very sought after, for instance, as back plates for power electronics[9]. The set of ordered metallic alloys is just a subset of possibly useful materials for making back plates. Other materials, such as carbides, nitrides and diamond can have the wanted high thermal conductivity, low thermal expansion and high stability. We have not considered such materials in the present study, as these materials often arrange themselves in much more complicated structures than the inter-metallic alloys, whereby it becomes computationally more demanding to make calculations of predictive quality.

It is a simple task to find the elementary metal in the database with the lowest compressibility. It is Os, which was recently shown both theoretically and experimentally[10] to be less compressible than even diamond at elevated pressures. Os is extremely expensive, however, and Os back plates on power electronics is not a viable option. The question is now how much one is willing to pay for a given level of compressibility. If there is a way of *a priori* answering this question, then one can effectively combine the two variables, price and compressibility into one variable, thereby reducing the dimensionality of the problem to an optimisation problem in one dimension. This closely resembles the approach by Ashby (see reference 1), who on the basis of engineering knowledge and dimensional analysis for many problems can reduce the initial variables to so-called materials indices, which can then be optimised. Sometimes, however, there is no *a priori* way of weighting the importance of the individual properties or materials indices against each other. This is an example of a multi-objective optimisation problem, which is often discussed in the economics literature, where one is trying to maximize the utility of a given investment while minimizing the price. There, a common solution is to use the Pareto-optimal set, as defined by the influential neo-classical economist Vilfredo Pareto in the beginning of the last century[11]. A discussion of Pareto-optimisation in more dimensions is given in reference 12. In multi-objective optimisation, there is normally not one single solution, which is optimal in all respects. Still some solutions are superior to others. These are called the non-dominated solutions and are defined as those for which it is impossible to improve one property, without making another property worse. The Pareto-optimal set is the set of all non-dominated solutions.

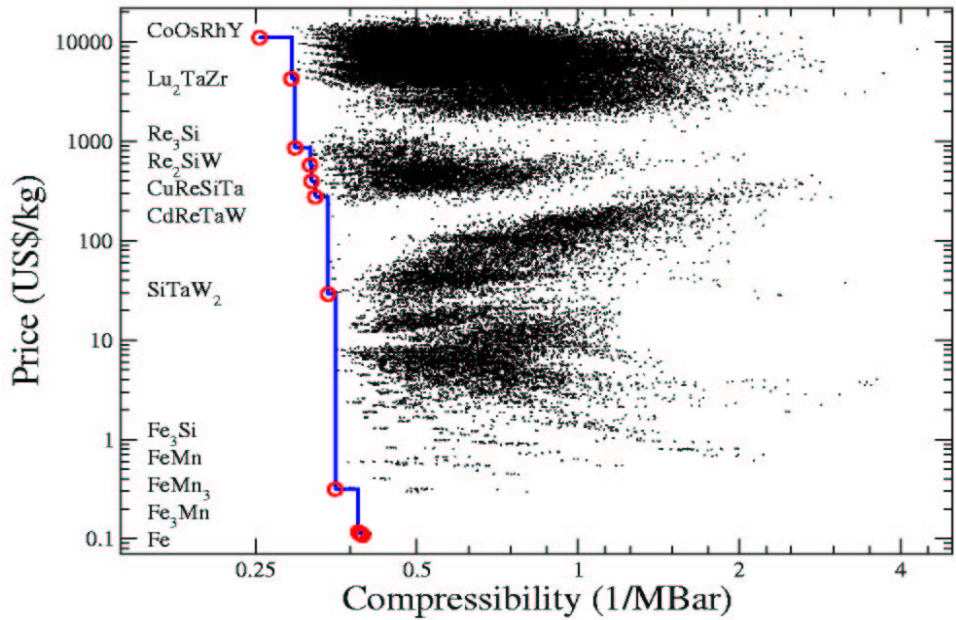

**Figure 1. Two-dimensional Pareto optimisation.** Theoretical compressibilities for 64,149 four-component alloys including the Pareto-optimal set of alloys with respect to low price and low compressibility.

In figure 1 we present the 64,149 alloys as small dots. They are positioned according to their compressibility along the x-axis and their price along the y-axis. The alloy prices are determined from the 1998 commodity prices of the pure elements[13] from which they are made. The cost of elemental extraction and purification is thus included, whereas the actual expenses in making the alloys afterwards are neglected, as these expenses would depend on the actual production process (method, quantity, rate, etc.). The alloys, which are Pareto-optimal with respect to low price and compressibility are connected by a solid line. This line, commonly called the Pareto-front, bounds the set of alloys 'from below' and therefore contains the most interesting solutions: given an alloy, which is not in the Pareto-optimal set it is always possible to find an alloy in the set, which is both cheaper and has lower compressibility. The Pareto-front in figure 1 immediately suggests why iron and iron alloys have found so widespread use for applications where high hardness (low compressibility) is desired at a reasonable price. The iron containing alloys are two orders of magnitude less expensive than the cheapest improvement – the tungsten and tantalum containing alloys. To decrease the compressibility even further, one has to add rhenium or eventually osmium, whereby the alloys become another one or two orders of magnitude more expensive, respectively.

In Figure 2 we show a Pareto-optimisation of the alloy database with respect to high stability as well as low price and low compressibility. In this figure all the dominated solutions have been removed and the non-dominated solutions are depicted as boxes. The corner of each box is placed at the point in a three-dimensional coordinate system, which corresponds to the formation energy, compressibility and price of that alloy. The boxes are coloured according to the qualities of the given alloy: The more negative the formation energy the more red the colour, the lower the compressibility the more green the colour, and the less expensive the more blue the colour. A table with names, formation energy, compressibility and price for each of the 82 Pareto-optimal alloys in figure 2 is available at http://www.fysik.dtu.dk/CAMPMDB/QuaternaryAlloys/. The Pareto optimal set as depicted in Figure 2 reproduces some well-known characteristics of alloys. Focussing on the regions where a single parameter is fully optimised we see that the most stable alloys are obtained by combining late and early transition metals like PtY or PtSc because the formation of a common half-filled d-band from an almost filled and an almost empty one gives the largest stability[14]. The alloys with the lowest compressibility typically contain Os or other transition metals from the central region of the periodic table where the compressibility of the pure elements is low again due to the electronic d-band, and the cheapest materials are as already mentioned not unexpectedly the iron-containing compounds. However, there are also interesting new and less obvious information in the data set. For example the strong presence of silicides (as for example RuSi) points to some interesting possibilities for mid-range cost materials with low compressibility and high stability in cases where the potential silicon oxidation problem can be controlled.

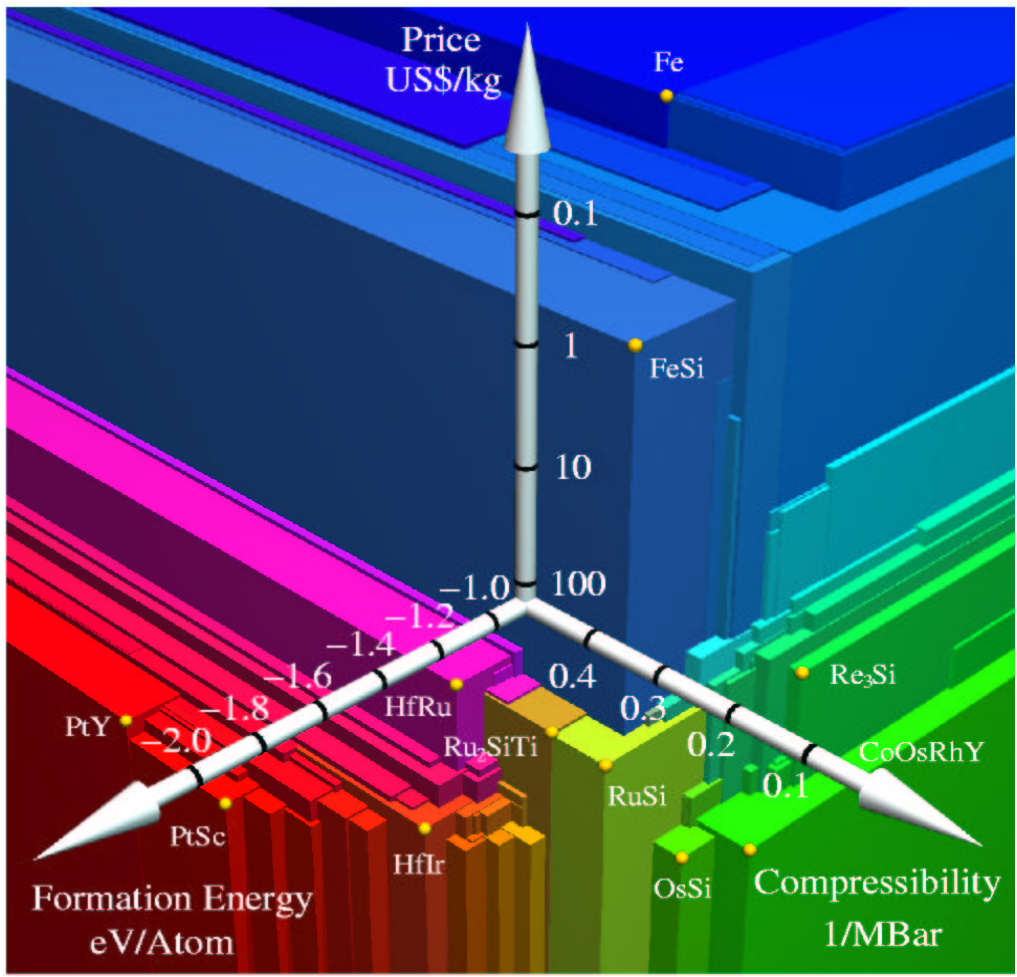

**Figure 2 Three-dimensional Pareto optimisation.** Pareto-optimal set of alloys with respect to high stability, low compressibility and low price.

Although finding the Pareto-optimal set in multi-objective optimisation of more than 3 properties does not pose any further challenge, it does become more difficult to depict the results graphically. When you only have a few materials to choose from, it might not make a big difference, how you pick the optimal ones. There will be very few, and you can have a good intuition about what to look for. When designing very specific materials on the other hand, which have to fulfil many constraints, and when you have maybe millions or billions of materials to chose between, it is going to make a very big difference which method you use to generate good compromises. We propose using Pareto-optimality as such a method in the future, where high-throughput experimental methods and much faster computers are going to vastly expand the number of new accessible materials.

We acknowledge support from Danish Center for Scientific Computing through grant no. HDW-1101-05